\documentclass[preprint,superscriptaddress,amsmath,amssymb,aps,prl]{revtex4-1}

\usepackage{graphicx}
\usepackage{times}
\usepackage{dcolumn}
\usepackage{bm}
\usepackage{amsmath}
\usepackage{amsfonts}
\usepackage{wasysym}
\usepackage{amssymb}
\usepackage{colortbl}
\usepackage{color}
\usepackage[version=3]{mhchem}
\usepackage{url}
\usepackage{hyperref}
\usepackage{pifont}

\begin{document}

\title{High pressure x-ray study of spin-Peierls physics in the quantum spin chain material TiOCl}

\author{Costel R. Rotundu}
\email{e-mail address: CostelRRotundu@gmail.com}
\affiliation{Stanford Institute for Materials and Energy Sciences, SLAC National Accelerator Laboratory, 2575 Sand Hill Road, Menlo Park, CA 94025, USA}
\author{Jiajia Wen}
\affiliation{Stanford Institute for Materials and Energy Sciences, SLAC National Accelerator Laboratory, 2575 Sand Hill Road, Menlo Park, CA 94025, USA}
\author{Wei He}
\affiliation{Stanford Institute for Materials and Energy Sciences, SLAC National Accelerator Laboratory, 2575 Sand Hill Road, Menlo Park, CA 94025, USA}
\affiliation{Department of Materials Science and Engineering, Stanford University, Stanford, CA 94305, USA}
\author{Yongseong Choi}
\affiliation{Advanced Photon Source, Argonne National Laboratory, Argonne, IL 60439, USA}
\author{Daniel Haskel}
\affiliation{Advanced Photon Source, Argonne National Laboratory, Argonne, IL 60439, USA}
\author{Young S. Lee}
\affiliation{Stanford Institute for Materials and Energy Sciences, SLAC National Accelerator Laboratory, 2575 Sand Hill Road, Menlo Park, CA 94025, USA}
\affiliation{Department of Applied Physics, Stanford University, Stanford, CA 94305, USA}

\maketitle
\section*{Abstract}

The application of pressure can induce transitions between unconventional quantum phases in correlated materials. The inorganic compound TiOCl, composed of chains of S=1/2 Ti ions, is an ideal realization of a spin-Peierls system with a relatively simple unit cell. At ambient pressure, it is an insulator due to strong electronic interactions (a Mott insulator). Its resistivity shows a sudden decrease with increasing pressure, indicating a transition to a more metallic state which may coincide with the emergence of charge density wave order. Therefore, high pressure studies of the structure with x-rays are crucial in determining the ground-state physics in this quantum magnet. In ambient pressure, TiOCl exhibits a transition to an incommensurate nearly dimerized state at $T_{c2}=92$~K and to a commensurate dimerized state at $T_{c1}=66$~K. Here, we discover a rich phase diagram as a function of temperature and pressure using x-ray diffraction on a single crystal in a diamond anvil cell down to $T=4$~K  and pressures up to 14.5 GPa. Remarkably, the magnetic interaction scale increases dramatically with increasing pressure, as indicated by the high onset temperature of the spin-Peierls phase. At $\sim$7 GPa, the extrapolated onset of the spin-Peierls phase occurs above $T=300$~K, indicating a quantum singlet state exists at room temperature. Further comparisons are made with the phase diagrams of related spin-Peierls systems that display metallicity and superconductivity under pressure.

\section*{Introduction}

Quantum magnets composed of interacting S=1/2 magnetic ions exhibit a wide variety of phenomena with phases like quantum spin liquids or valence bond crystals~\cite{Anderson1973}. The application of pressure is a clean way to continuously adjust the interaction parameters to explore the ground state physics. For example, a continuous phase transition as a function of pressure at $T\rightarrow$0 would be indicative of a quantum phase transition. In one-dimensional systems, a fascinating ground state is the spin-Peierls state~\cite{Pytte1974}, arising from the coupling of S=1/2 spins to the lattice in Heisenberg antiferromagnetic spin chain materials. A periodic deformation of the lattice (dimerization) along the chain takes place below a characteristic temperature $T_{SP}$. The deformation of the lattice enhances the exchange interaction between neighboring magnetic atoms that causes formation of singlet pairs of localized electrons (also known as a valence bond crystal). This phase is robust, occurring in a minimal model involving nearest neighbor magnetic exchange $J$ and a coupling to thee-dimensional phonons. The spin-Peierls transition has been observed in many organic compounds~\cite{Bray1975,Jacobs1976,Huizinga1979} but only in two inorganic systems, CuGeO$_{3}$~\cite{Hase1993} and TiOCl~\cite{Seidel2003,Abel2007}. However, CuGeO$_{3}$ is not an ideal realization of a spin-Peierls system due to the presence of significant next nearest neighbor exchange coupling along the chain~\cite{Riera1995}. TiOCl is an ideal realization which has been shown to undergo the canonical soft phonon transition to the spin-Peierls state.\cite{Abel2007} TiOCl crystalizes in an orthorhombic FeOCl-type structure in the space group $Pmmm$ with buckled Ti-O bi-layers in $ab$ plane separated by Cl layers (Fig. 1a). The Ti$^{3+}$ ions, in the 3d$^{1}$ electronic configuration, have an orbital arrangement forming quasi-1D S=1/2 Heisenberg spin chains along the crystallographic $b$-direction with a nearest neighbor magnetic exchange of $J\approx$660K~\cite{Seidel2003}. Adjacent chains are displaced by $b/2$ which leads to frustration of the interchain magnetic interaction, resulting in effectively decoupled spin chains. Here, we investigate the phase transitions of TiOCl as a function of pressure at low temperatures. We find that quantum effects are surprisingly robust, indicating singlet formation as high as room temperature under pressure.

TiOCl exhibits two phase transitions at $T_{c2}\approx92$~K and $T_{c1}\approx66$~K at ambient pressure, corresponding to an incommensurately modulated~\cite{Shaz2005,Abel2007} and a commensurate state~\cite{Ruckamp2005,Krimmel2006,Schonleber2006,Abel2007}, respectively. At high temperatures, $T>T_{c2}$, the material is a paramagnetic Mott insulator and at low temperatures, $T<T_{c1}$, the system is in a dimerized singlet state (Fig. 1b). Below $T_{c2}$, the chains are dimerized with periodic discommensurations, and inelastic x-ray scattering of the soft phonon indicates that $T_{c2}$ corresponds to the spin-Peierls transition temperature.\cite{Abel2007} Below $\approx150$~K a pseudo-gap in the spin excitations has been reported~\cite{Imai2003,Caimi2004,Lemmens2004,Hemberger2005,Clancy2007}. Interestingly, similar gaps in the magnetic excitations of Mott-insulators often appears in S=1/2 quantum magnets, including the high-$T_C$ cuprates. Calculations have suggested that TiOCl may be in the proximity of an insulator-to-metal transition, and further, it may become superconducting with doping~\cite{Beynon1993,Seidel2003,Craco2006,Kuntscher2010,Zhang2010} or application of pressure. The limited elemental doping performed to date on TiOCl showed no metallization. Two proposed paths not yet explored are doping of alkali metals together with organic ligands or application of pressure on electron-doped TiOCl~\cite{Zhang2010}. Application of high pressure on undoped TiOCl may be a promising route. An example of a spin-Peierls system exhibiting pressure-induced superconductivity is the organic compound (TMTTF)$_{2}$PF$_{6}$~\cite{Adachi2000,Jaccard2001}.

\begin{center}
\begin{figure}
\begin{center}
\includegraphics[width = 0.8\textwidth]{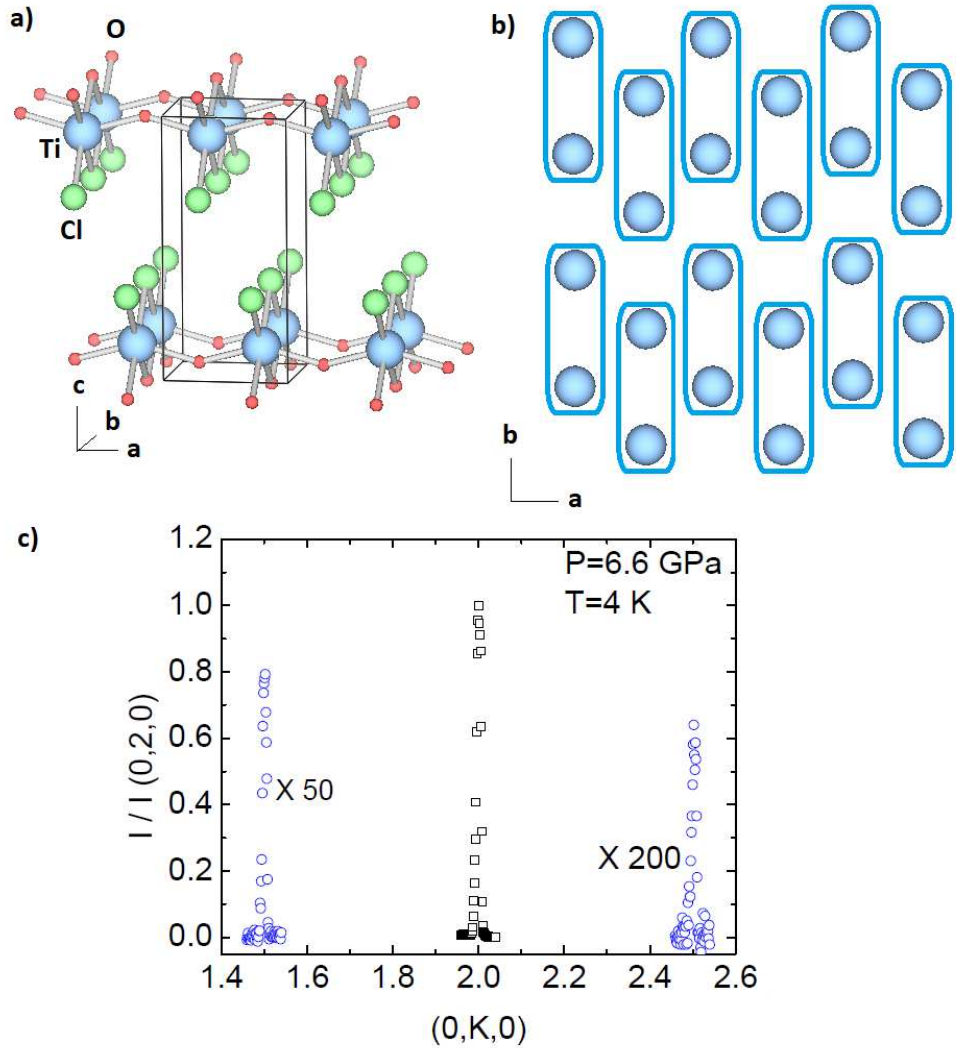}
\caption{\textbf{Structure of TiOCl}. (\textbf{a}) Crystal structure of TiOCl. (\textbf{b}) Representation of the commensurate dimer singlet state at $T<T_{c1}$ (the spin-Peierls state). (\textbf{c}) Scans along K through the commensurate (0,1.5,0) and (0,2.5,0) peak positions showing the peak intensity normalized to the nearest Bragg peak (0,2,0). For clarity, the superlattice peaks were magnified by a factor of 50 and 200, respectively.}
\label{fig:Fig1}
\end{center}
\end{figure}
\end{center}

Optical measurements in the visible and infrared spectra suggest that the system becomes more metallic as the pressure is increased above $\sim$13 GPa~\cite{Kuntscher2006,Kuntscher2007,Kuntscher2008,Kuntscher2010}. Direct electrical resistivity measurements up to 24 GPa confirmed that the overall resistivity decreases about seven orders of magnitude~\cite{Forthaus2008} to a weaker insulator, but true metallization remains elusive. Early x-ray powder diffraction experiments revealed an orthorhombic $Pmmn$ to monoclinic $P2_{1}/m$ structural phase transition at 16 GPa in TiOCl~\cite{Kuntscher2008}, at room temperature. Later, it was found that aside from this structural transition at $\approx$15 GPa~\cite{Ebad2010,Kuntscher2010}, there may be another structural transition at $\approx$22 GPa~\cite{Ebad2010,Kuntscher2010}. All the aforementioned structural experiments were performed on powders, where He gas was used as the pressure medium, with the exception of the experiments of Blanco-Casona $\emph{et al.}$~\cite{Blanco2009} where a CH$_{3}$OH:C$_{2}$H$_{5}$OH 4:1 pressure medium was used and the low pressure transition was found at $\approx$10 GPa. Prodi and coworkers studied the low temperature ($T=6$~K) single crystal diffraction found that the structural transition occurs at 13.1 GPa~\cite{Prodi2010} when methanol:ethanol 4:1 was used as pressure medium. The transition was accompanied by an incommensurate charge-density wave perpendicular to the original spin-chain direction ($b$ axis)~\cite{Prodi2010}. This pressure induced charge order is intriguing, which suggests that the more metallic state at high pressure has a Fermi surface which is gapped due to charge density wave order. As mentioned earlier, the weakening of the insulating behavior by over seven orders of magnitude gives hope that doping combined with the application of pressure may result in metallicity and perhaps even superconductivity. Therefore, further investigation of the charge-order and a full study of the temperature-pressure phase diagram has been performed to better understand this fascinating system. This was done with synchrotron x-ray diffraction on a single crystal in a high pressure diamond anvil cell down to $T=4$~K and in pressures up to approximately 14.5 GPa.

\section*{Results}

For the low temperature single crystal x-ray scattering measurements in a diamond anvil cell (DAC), the pressure medium was selected to provide hydrostatic pressure, yet remain solid in the parameter space of interest so that the sample does not lose alignment. Here, we used a methanol:ethanol 4:1 mixture as the pressure medium. A helium-membrane-controlled diamond anvil cell was used to allow the sample pressure to be changed in situ at base temperature with better than 0.05 GPa resolution. Details on the sample environment are given in the $\emph{Methods}$ section. For mapping the complete T -- P phase diagram for pressures up to 14.5 GPa and temperatures down to 4 K, we organized the experiment into three sets of measurements. The DAC was loaded and then compressed to about 0.3 GPa at room temperature, then cooled to 100 K and finally compressed further to 6 GPa before cooling further to the base temperature of 4 K. The first set of measurements was carried out at a nominal pressure of 6 GPa at various temperatures on warming, within the range of $T=4 - 215$~K. The second set of measurements was carried at fixed temperature of 182~K and varying the pressure in the range 8 - 12 GPa. The purpose of these two sets of measurements is to determine the high temperature boundary of the phase diagram. A third set of measurements was performed at nominal pressures of 12 and 15 GPa, for several temperatures in the 4 - 182~K and 4 - 210~K range, respectively, with the aim of determining the high pressure boundary of the phase diagram. The trajectories through the phase diagram were chosen to keep the pressure medium solid and hence maintain crystal alignment.

\begin{center}
\begin{figure}
\begin{center}
\includegraphics[width = 0.8\textwidth]{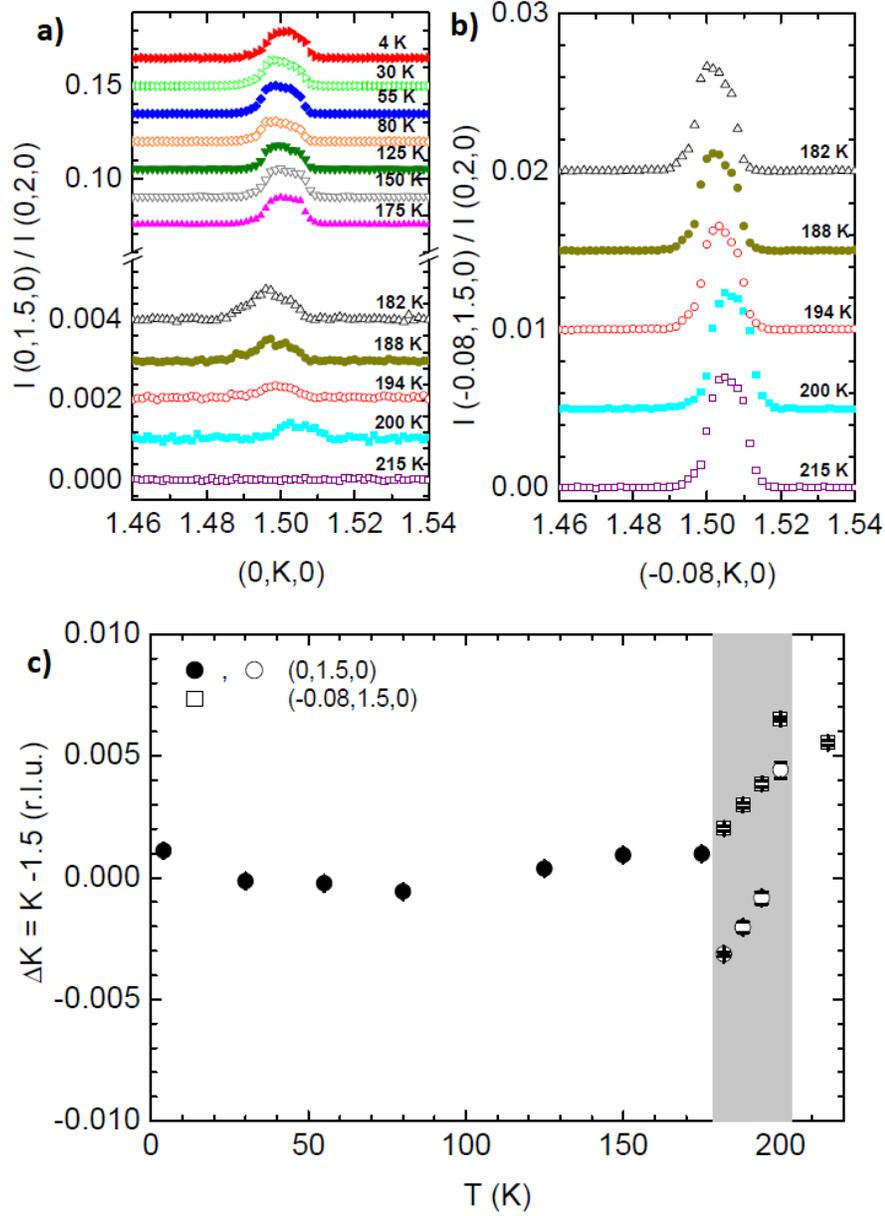}
\caption{\textbf{Temperature dependence of the superlattice reflections (H,1.5,0) at nominal 6 GPa: Scans along K}. (\textbf{a}) Scans along K through the (0,1.5,0) peak position at nominal pressure 6 GPa for temperatures between 4 and 215~K. (\textbf{b}) Scans along K through the (-0.08,1.5,0) peak position for temperatures between 182 and 215~K. (\textbf{c}) $\Delta$K=K -- 1.5 (r.l.u.) vs. $T$ for both (0,1.5,0) and (-0.08,1.5,0) peaks. The grey area marks the temperature interval of coexistence of both commensurate (0,1.5,0) and incommensurate (-0.08,1.5,0) peaks. The error bars are smaller than the size of the symbols.}
\label{fig:Fig2}
\end{center}
\end{figure}
\end{center}

\begin{center}
\begin{figure}
\begin{center}
\includegraphics[width = 0.8\textwidth]{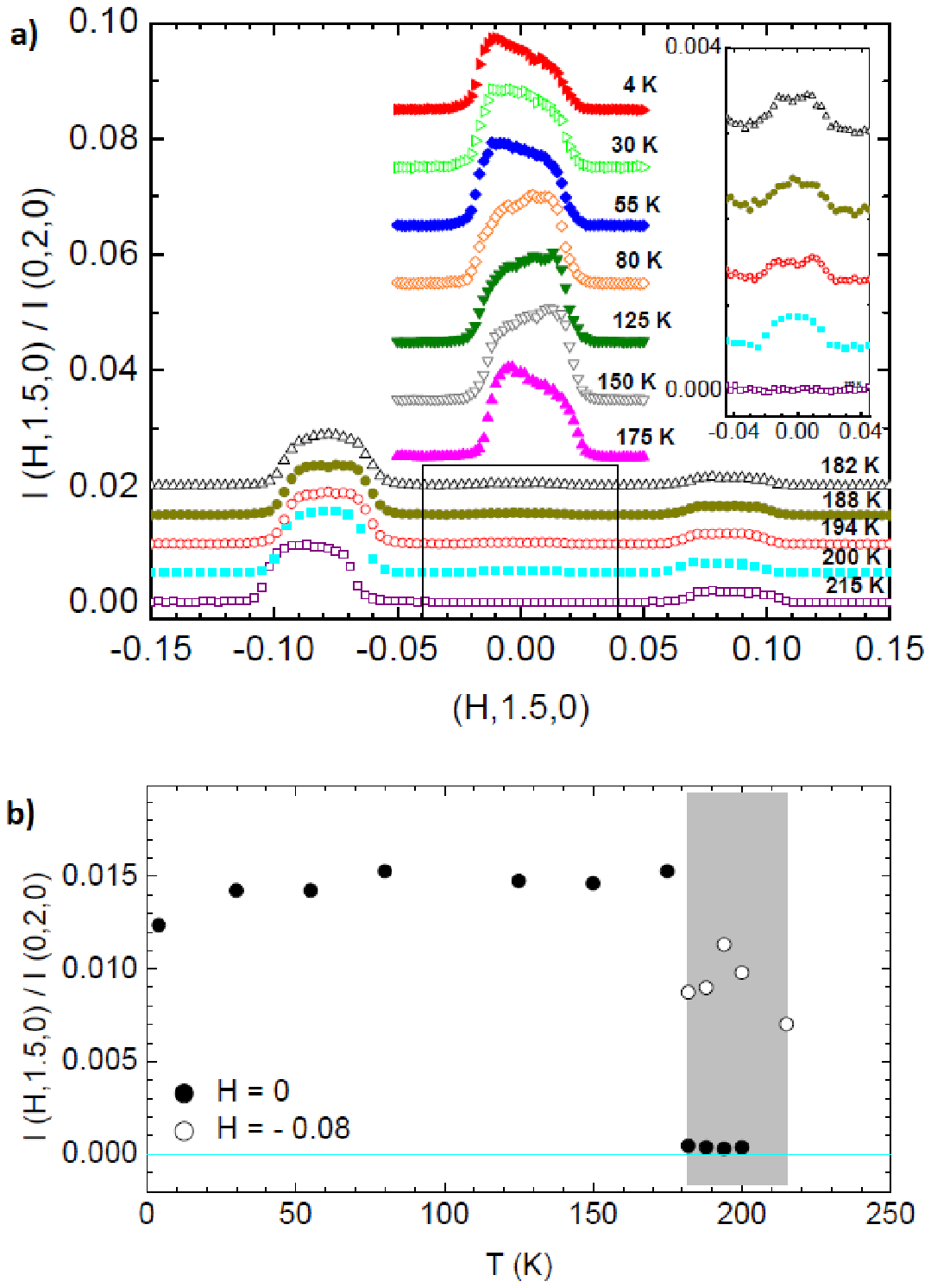}
\caption{\textbf{Temperature dependence of the superlattice reflections (H,1.5,0) at nominal 6 GPa: H-scans}. (\textbf{a}) Scans along (H,1.5,0) direction at nominal pressure 6 GPa for temperatures between 4 and 215~K. Below 175~K there is only the commensurate peak at H=0. Between 200 and 182~K both the commensurate (0,1.5,0) and incommensurate ($\pm$0.08,1.5,0) peaks are present. At 215~K only the incommensurate ($\pm$0.08,1.5,0) peaks are present. The right window is a magnification of the data in the black rectangle centered at H=0. (\textbf{b}) Integrated intensity of the superlattice peaks as a function of temperature.}
\label{fig:Fig3}
\end{center}
\end{figure}
\end{center}

For the first set of measurements at a nominal pressure of 6 GPa, the differential thermal expansion of the gasket, diamonds and the brass case of the cell causes the actual pressure inside the sample chamber to change slightly with temperature even if the pressure of the driving membrane mechanism is kept constant. The actual pressure at each temperature was measured, and typical values were 6.58 GPa at $T=4$~K and 7.7 GPa at $T=215$~K. Scans along the K direction through the commensurate (0,1.5,0) and (0,2.5,0) superlattice peak positions (normalized to the Bragg peak intensity at (0,2,0)) are shown in Fig.1c. For clarity, the superlattice peaks were magnified by 50 and 200, respectively. The temperature dependence of the dimerization is shown in Figure 2a, where K-scans are plotted through the commensurate (0,1.5,0) peaks at 6 GPa for temperatures between 4 and 215 K. Above 182~K a new set of incommensurate peaks appear near positions (-0.08,1.5,0) and (0.08,1.5,0). It should be noted the sudden drop in intensity (by a factor of 40) of the commensurate peak at 182~K, coincides with the emergence of the incommensurate ($\pm$0.08,1.5,0) peaks. The K-scans through the incommensurate (-0.08,1.5,0) peak position at nominal pressure of 6 GPa for temperatures between 182 and 215~K are shown in Fig. 2b. The $\Delta$K=K -- 1.5(r.l.u.) component of the displacement wave vector versus temperature for both (0,1.5,0) and (-0.08,1.5,0) peaks is shown in Fig. 2c. The K position of the peaks was determined from Gaussian fits and the error bars are less or equal to symbols size. The grey area marks the temperature interval of coexistence of both commensurate (0,1.5,0) and incommensurate (-0.08,1.5,0) peaks (we will refer to as the mixed phase). One other observation is that at 6 GPa the $\Delta$K component of the displacement wave vector of the incommensurate peak is several fold smaller compared to that at ambient pressure~\cite{Abel2007}.

The intensity of the incommensurate peak is about one order of magnitude smaller than that of the commensurate peak. Figure 3a shows H-scans through the same superlattice peaks normalized to the nearby (0,2,0) Bragg peak, at nominal 6 GPa for temperatures between 4 and 215~K. At 182~K, incommensurate peaks appear near positions (-0.08,1.5,0) and (0.8,1.5,0). In the temperature interval 182 - 200~K the commensurate and incommensurate peaks coexist. The inset of Fig. 3a shows the area within the rectangle centered at H=0 to more clearly depict the commensurate (0,1.5,0) peak intensities for temperatures 182~K to 215~K. At 215~K, the commensurate peak is entirely suppressed leaving only the two incommensurate peaks. The lower panel shows normalized intensity of (H,1.5,0) peaks vs. temperature. The intensity of the commensurate (0,1.5,0) peak drops rapidly when the system enters the mixed phase, consistent with the same observation from the K-scans (Fig. 2a). This is also consistent with the observed rapid drop in intensity at $T_{c1}$ in ambient pressure conditions~\cite{Abel2007}).

The incommensurate modulation along the $a$-direction continues to evolve as a function of pressure. Figure 4a shows H-scans at $T=182$~K for several pressures in the 8 - 12 GPa range. This set of scans provide information on the high temperature boundary of the T -- P phase diagram. Within this pressure range, the H scans show both commensurate (0,1.5,0) and incommensurate ($\approx\pm$0.08,1.5,0) peaks up to 11.5 GPa. At 12 GPa the commensurate peak is suppressed leaving only the incommensurate peaks. In Figure 3b is plotted $\Delta$H=(H$_{+}$ -- H$_{-}$)/2 versus pressure at 182~K. H$_{+}$ and H$_{-}$ determined from fits of the peaks with Gaussians. The error bars in Fig. 4b are smaller or equal to the size of symbols. With application of pressure, the incommensurability rapidly increases. At this temperature (182~K), pressure has a rather strong effect on the incommensurability of the ($\approx\pm$0.08,1.5,0) superlattice peak.

\begin{center}
\begin{figure}
\begin{center}
\includegraphics[width = 0.8\textwidth]{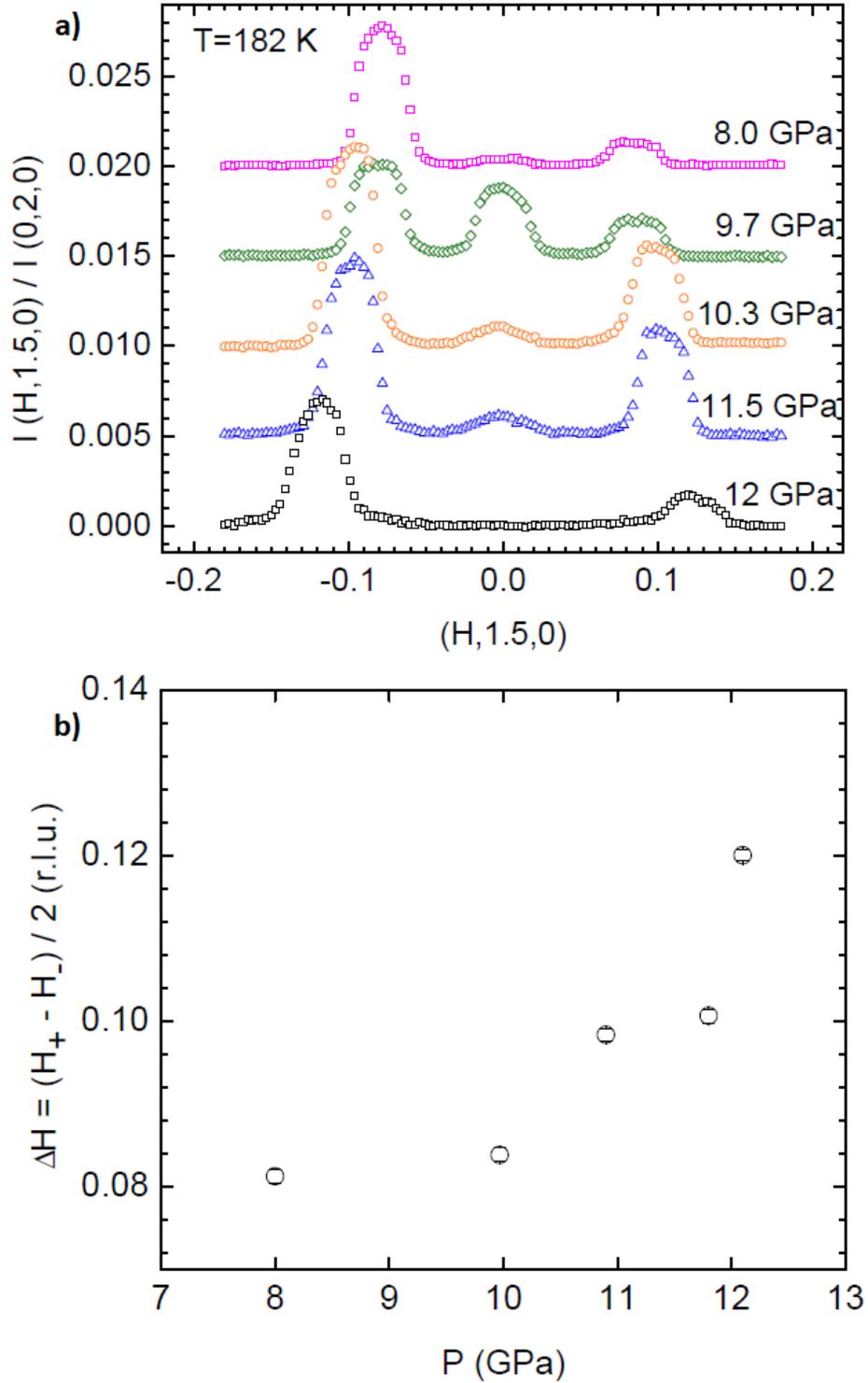}
\caption{\textbf{Pressure dependence of the superlattice reflections (H,1.5,0) at 182~K}. (\textbf{a}) (H,1.5,0) scans at 182~K for different pressures. (\textbf{b}) The incommensurate wavector (relative to the commensurate dimerized superlattice position)  as a function of pressure at $T=182$~K.}
\label{fig:Fig4}
\end{center}
\end{figure}
\end{center}

\begin{center}
\begin{figure}
\begin{center}
\includegraphics[width = 0.8\textwidth]{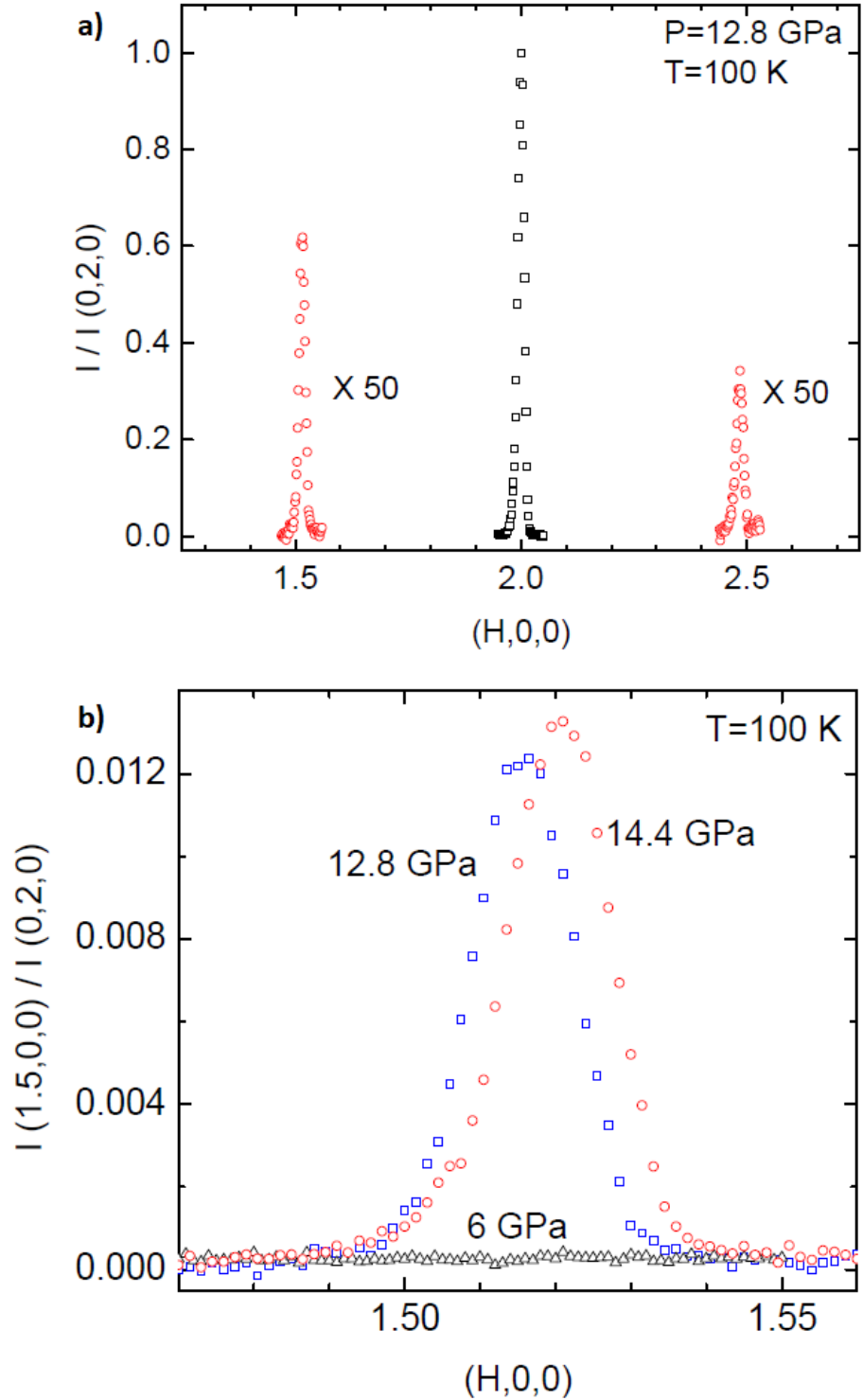}
\caption{\textbf{The high-pressure incommensurate $a$-axis modulation reflections near (1.5,0,0) and (2.5,0,0)}. (\textbf{a}) Scans along K through the (1.5,0,0) and (2.5,0,0) peak positions showing the intensity normalized to the nearby (2,0,0) Bragg peak, at $P=12.8$ GPa and $T=100$ K. (\textbf{b}) H-scans through the (1.5,0,0) peak at 100~K for $P=12.8$~and 14.4 GPa. At 6 GPa, there is no discernable $a$-axis modulation peak.}
\label{fig:Fig5}
\end{center}
\end{figure}
\end{center}

\begin{center}
\begin{figure}
\begin{center}
\includegraphics[width = 0.9\textwidth]{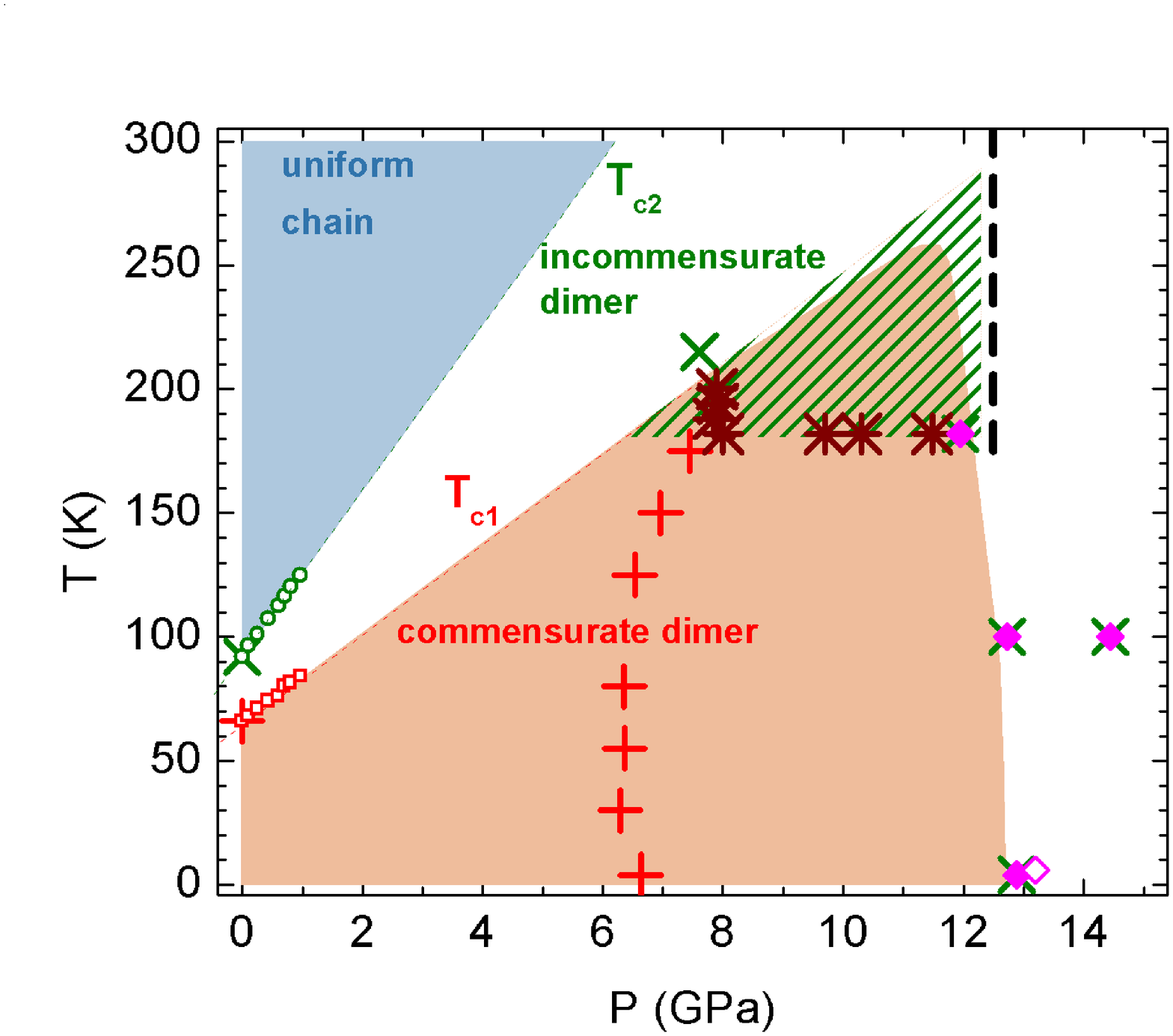}
\caption{\textbf{T(K) -- P(GPa) phase diagram of TiOCl}. ``$+$'' mark the commensurate peaks (commensurate dimmer phase), ``$\times$'' the incommensurate peaks (incommensurate dimmer phase), and ``$\varhexstar$'' marks the ``mixed'' phase (coexistence of both commensurate and incommensurate phases), \ding{117} marks the incommensurate $a$-axis modulation peak and $\Diamond$ marks the original $a$-axis modulation peak from Prodi $\emph{et al.}$~\cite{Prodi2010}, the {\scriptsize $\square$} and $\circ$ mark the two transitions $T_{c1}$ and $T_{c2}$ from the corresponding peaks in d$\chi$(T)/dT, adapted from Blanco-Canosa $\emph{et al.}$~\cite{Blanco2009}. The vertical dashed line marks the resistivity crossover~\cite{Forthaus2008,Kuntscher2010}, in vicinity of the orthorhombic $Pmmn$ to monoclinic $P2_{1}/m$ structural phase transition (see text).}
\label{fig:Fig6}
\end{center}
\end{figure}
\end{center}

Previous x-ray studies by Prodi $\emph{et al.}$~\cite{Prodi2010} revealed the emergence of a new ground state for pressures above $\sim13$ GPa, characterized by a pair of incommensurate superlattice peaks with modulation vector $\delta_a=0.48$ along the $a$-axis (which is perpendicular to the low-pressure spin chain direction). These peaks  arise above the critical pressure to the more metallic state, and may be related to charge density wave ordering in a more two-dimensional electronic system. We observed similar $a$-axis modulations and find that this new phase exists over an extended pressure and temperature range. Figure 5 shows the incommensurate $a$-axis modulation reflections at (1.52,0,0) and (2.48,0,0). Figure 5a shows longitudinal scans through the peaks normalized to the nearby (2,0,0) Bragg peak intensity, at $P=12.8$ GPa and $T=100$~K. Figure 5b shows longitudinal scans as a function of pressure at 100~K for $P=12.8$ and 14.4 GPa. As expected, the $a$-axis modulation disappears at 6 GPa and $T=100$~K, which is well below the critical pressure.

\section*{Discussion}
Our observations of weak x-ray superlattice peaks in high pressures shed light on two important aspects of the spin-Peierls physics in TiOCl. First, the application of pressure dramatically enhances the magnetic energy scale, and hence the critical phase boundary for the dimerization transition. Second, near the critical pressure of $\sim13$~GPa, the commensurate superlattice peaks disappear; however, incommensurate modulations of the structure remain. In fact, the modulations are characterized by two types of discommensurations: around both the (0, $2\pm 0.5$, 0) and ($2\pm 0.5$, 0, 0) peaks (using the low temperature orthorhombic notation). Our main results can be summarized in the phase diagram depicted in Fig. 6. It had been hypothesized from high pressure magnetization measurements that, if the increase rate of the transitions remain constant ($\partial$lnT$_{c1}$/$\partial$P)=2.88$\times$10$^{-1}$ K/GPa and ($\partial$lnT$_{c2}$/$\partial$P)=3.4$\times$10$^{-1}$ K/GPa, the spin-Peierls transition should reach room temperature at a pressure somewhere around 6 GPa~\cite{Blanco2009}. Our high temperature x-ray scattering data provide unique evidence on the dimerization transition temperature in high pressures. We find that $T_{c1}$ appears to increase almost linearly with pressure up to at least 7 GPa. Therefore TiOCl under pressure ($>$7 GPa) is likely the first example of a room temperature quantum singlet state. This is facilitated by the large exchange constant $J\approx660$~K~\cite{Seidel2003,Kataev2003} at ambient pressure, the highest reported among spin-Peierls compounds. At a pressure of 9 GPa the direct exchange interaction $J$ is estimated to be 3300~K, which would be the highest known exchange coupling in a quantum magnet (for instance in La$_{2}$CuO$_{4}$ $J\approx1500$~K and in Sr$_2$CuO$_3$ $J\approx2800$~K~\cite{Igor2009} in ambient pressure).

We can compare the phase diagrams of TiOCl with the phase diagram of another spin-Peierls system, the organic (TMTTF)$_{2}$PF$_{6}$, which becomes superconducting at 4 GPa (up to 7 GPa)~\cite{Adachi2000,Jaccard2001}. The phase diagram of (TMTTF)$_{2}$PF$_{6}$ is in fact more general, being common to Fabre and Bechgaard salts~\cite{Jaccard2001,Yu2004,Dressel2007} and to the organic superconductors~\cite{Jerome1991}. With application of pressure, in both (TMTTF)$_{2}$PF$_{6}$ and TiOCl, density wave order emerges at higher pressures beyond the spin-Peierls state. The pressure-induced metallic phase present in (TMTTF)$_{2}$PF$_{6}$ was not yet reached in TiOCl, but, within the highest pressure measured to date (24 GPa), a large $10^7$-fold decrease of the overall electrical resistivity has been reported, as mentioned earlier~\cite{Forthaus2008}. The fastest decrease in overall resistivity with application of pressure takes place for pressures up to 13 GPa, critical pressure coinciding with fastest decrease of the activation energy~\cite{Forthaus2008}. This pressure of the resistivity crossover (between insulator and weak insulator) is marked in Fig. 6 by the vertical dashed line. Therefore, the similarity in the phase diagram of the two spin-Peierls systems is another hint of possible metallization at very high pressures. The relatively higher critical pressures of the different ground states in TiOCl compared with those in (TMTTF)$_{2}$PF$_{6}$ may be related to the overall larger energy scale $J$ of TiOCl compared to the other spin-Peierls systems. The pressure medium (steatite) used in the (TMTTF)$_{2}$PF$_{6}$ experiments~\cite{Jaccard2001} (as discussed in the $\emph{Methods}$ section) is less hydrostatic, and it may be that uniaxial forces play an vital role in the transitions. It is known for instance that the Temperature-Strain phase diagram of organic $\alpha$-(BEDT-TTF)$_{2}$I$_{3}$ displays superconductivity only with the strain along $a$-axis~\cite{Tajima2002}. Similar strain studies on TiOCl should also be pursued.

In this paper, using synchrotron x-ray scattering on a TiOCl single crystal in a diamond anvil cell we determined the phase diagram for this spin-Peierls system down to $T=4$~K and in pressures up to 14.5 GPa. Scans through the commensurate (0,1.5,0) position at a pressure of $\sim$7 GPa revealed incommensurate peaks persisting to the highest measured temperatures.  These measurements reveal a region of coexistence in the phase diagram. At higher pressures (above around 12 GPa) the commensurate peak is suppressed, leaving only the incommensurate modulation. Further, the incommensurate $a$-axis modulation peaks at (1.52,0,0) and (2.48,0,0) reveals a charge density wave phase that exists over an extended region of the phase diagram. This region of the phase diagram corresponds to the more metallic phase of TiOCl. The dramatic increase of the spin-Peierls temperature with pressure indicates that TiOCl appears to be the an example of a quantum singlet ground-state at room temperature. To further test the applicability of the soft phonon mechanism of the transition, future measurements of the phonons and spin excitations in comparable pressures would be most instructive.

\section*{Methods}

\textbf{Materials and crystal growth.} The single crystals of TiOCl were grown by the chemical vapor transport method from TiCl$_{3}$ and TiO$_{2}$ in a closed quartz tube into a temperature gradient~\cite{Schafer1958,Seidel2003}. Thermal convection caused by the temperature gradient transports the material, with single crystals growing at the colder end of the tube. Typical dimensions of the grown crystals were a few mm$^2$ in the $ab$ plane and 10-100~${\mu}$m along $c$ axis (stacking direction). The purity of the crystals was checked by x-ray powder diffraction and the orientation was confirmed using x-ray Laue. The thickness of the as grown crystal makes it ideal for the x-ray measurements without the need of cleaving or polishing. More in-depth characterization of the crystals was reported in earlier studies~\cite{Abel2007,Prodi2010}.

\textbf{Bulk property measurements.} Magnetization measurements were collected within a Quantum Design SQUID MPMS magnetometer. Data showing magnetic susceptibility versus temperature is shown in the Supplementary Fig. S1. Because of the low signal the measurement was carried in a moderate high magnetic field of 1 T (parallel with the crystallographic $ab$ plane of the sample) on several coaligned crystals of total mass of 2.5~mg. The very large spin coupling constants makes the two transitions insensitive to fields in this range.

\begin{center}
\begin{figure}
\begin{center}
\includegraphics[width = 0.8\textwidth]{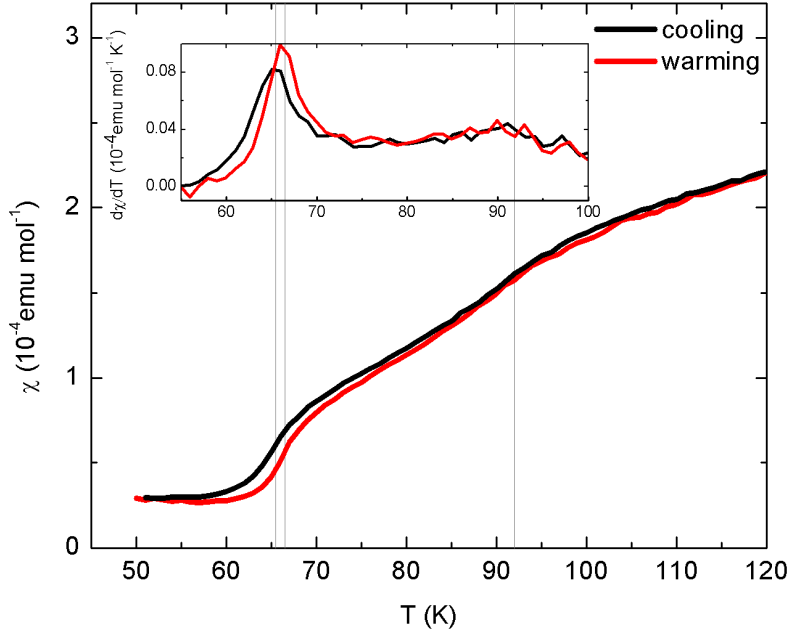}
\caption{\textbf{Supplementary Figure 1}. Magnetic susceptibility as a function of temperature of TiOCl single crystal in an applied magnetic field of $H=1$ T parallel with the ab plane. The grey lines correspond to $T_{c2}=92$~K when the system undergoes, upon cooling, to an incommensurate nearly dimerized state, and to $T_{c1}=66$~K when the system undergo to a commensurate dimerized state. First it is a second order transition and last is a first order transition, based on the presence of the hysteresis.}
\label{fig:FigS1}
\end{center}
\end{figure}
\end{center}

\textbf{High pressure x-ray diamond anvil cell.} At 20 keV incident photon energy, the diamonds are nearly transparent, making it ideal for high pressure x-ray scattering measurements. For the experiments, a Merrill-Bassett diamond anvil cell (DAC) consisting of two opposed 800~${\mu}$m culet diamonds was used. ``301'' stainless-steel with initial thickness of 250~${\mu}$m was indented first between the two diamonds to about 140~${\mu}$m thickness, and a centered 395~${\mu}$m hole was drilled. After, the gasket was placed on top of one of the anvils matching the indentation, the sample and a 30~${\mu}$m square piece of 5~${\mu}$m thick silver foil were placed inside the drilled hole (sample chamber) and finally the hole was filled with 4:1 methanol:ethanol mixture that served as pressure transmitting medium. For the TiOCl sample, a crystal 40~${\mu}$m~$\times$~60~${\mu}$m was cut from a $\approx$20~${\mu}$m thick larger crystal. When loaded, the $ab$ plane of the sample was parallel with the diamond's culets. A Ruby chip was placed in the chamber for loading pressure measurement. The expected change in thickness of the indented gasket filled with 4:1 methanol:ethanol pressure medium at 50 GPa and 5~K is from 140 thickness down to 50~${\mu}$m. The properties of the pressure-transmitting medium are discussed in the next section. Pressure was calibrated in situ against the lattice constant of silver, determined from position of three Bragg peaks [(200), (020), and (201)] of the silver foil inside the pressure chamber. A He-gas-driven membrane system was the generator of the applied force on the diamonds. The diamond-anvil cell was mounted on the cold finger of a closed-cycle cryostat with base temperature of about 4~K.

\textbf{Pressure-transmitting medium.} The closest to a perfect hydrostatic pressure medium up to high pressures is He gas. X-ray measurements at room temperature and 77~K up to 10 GPa showed also almost no difference in pressure gradients between He gas and 4:1 methanol:ethanol mixture pressure media~\cite{Tateiwa2009}. Still, x-ray scattering measurements (which are the most relevant to our study) at 5~K and for pressures up to 20 GPa show equal level of pressure inhomogeneity with $\Delta$P/P per-unit-area of $\pm$1.8 $\%$/(10$^4$ $\mu$m$^2$)~\cite{Yejun2010}. While He gas is characterized by a constant deviatoric stress of 0.021$\pm$0.011 GPa up to 16 GPa, 4:1 methanol:ethanol shows same level of anisotropy up to 10 GPa, which slightly increases up to 16 GPa. Therefore, for low temperature single crystal x-ray diffraction, He and 4:1 methanol:ethanol are considered the best pressure media. The main disadvantage of He gas is its high compressibility which causes the sample chamber to shrink significantly on gas loading under pressure. This reduces the sample to pressure medium volume ratio. One other disadvantage of use of He gas is, because of its diffusion, generation of cracks and eventual failures in the diamonds~\cite{Dewaele2006} (at Mbar pressures). But the choice of one over the other pressure medium in present experiment was dictated by the pressure range of measurements and by the pressure-chamber-to-sample volume ratio~\cite{Yejun2010}, making 4:1 methanol:ethanol mixture a better choice for the 10 GPa pressure range. For the particular system TiOCl, transmission measurements using 4:1 methanol:ethanol mixture and argon gas (another close to hydrostatic pressure medium similar to He gas) showed results that qualitatively agree~\cite{Kuntscher2006}. As a contrasting example, when CsI powder was used as pressure medium, the pressure-induced structural effects occurred at lower pressure (with approximately 4 GPa), due to less hydrostatic conditions~\cite{Kuntscher2006}. Blanco-Casona $\emph{et al.}$~\cite{Blanco2009} used the less hydrostatic pressure medium CH$_{3}$OH:C$_{2}$H$_{5}$OH 4:1 for their x-ray scattering experiments and found a lower pressure for the $Pmmn$ to $P2_{1}/m$ structural transition, namely $\approx$10 GPa.

\textbf{X-ray diffraction measurements.} Synchrotron x-ray diffraction experiments at high pressures and low temperatures were performed at the 4-ID-D beamline of the Advanced Photon Source at Argonne National Laboratory. The DAC was cooled in a $^4$He closed-cycle cryostat that was installed on the sample stage of a Huber psi-circle diffractometer. The experiment was in transmission (Laue) geometry, where x-ray beam traverses through the sample thickness and diffraction takes place in the vertical scattering plane. An incident x-ray energy of 20 keV was selected with a double-bounce Si(1 1 1) monochromator. A toroidal Pd mirror focused the beam to approx 100 $\times$ 200 $\mu$m$^2$ (VXH). The beam size was further reduced with a slit to match the sample size. To maximize on the ${\bf Q}$ resolution and eliminate uncertainties related to both the zero position and the overall scale of 2$\theta$ (which are issues commonly encountered with image-plate) a NaI point detector was used. The sample - detector distance was 1000 mm, and the detector slit size was 5 mm (horizontal) $\times$ 1 mm (vertical), giving a typical resolution of the instrument of less than $\Delta$($\mathbf{Q}$) $\approx$ 4 $\times$ 10$^{-3}$~${\rm \AA}^{-1}$~\cite{Yejun2010}. All superlattice peaks were resolution limited at all pressures and temperatures along H and K.

\section*{Acknowledgements}
The work at SLAC was supported by the U.S. Department of Energy (DOE), Office of Science, Basic Energy Sciences, Materials Sciences and Engineering Division, under Contract No. DE- AC02-76SF00515 (sample preparation, x-ray scattering, and data analysis). The use of the Advanced Photon Source, Argonne National Laboratory was sponsored by the Scientific User Facilities Division, Office of Basic Energy Sciences, US DOE.

\section*{Author contributions}
C.R.R. and Y.S.L. conceived the experiment. C.R.R., J.W., W.H., and Y.S.L. performed the measurements. Y.F., Y.C. and D.H. prepared the beamline instrument for experiment. C.R.R. analyzed the data and wrote the manuscript with input from all authors. The authors thank Yejun Feng for assistance with the experiment and Hongchen Jiang for useful discussions.

\section*{Additional information}
\textbf{Competing financial interests:} The authors declare no competing financial interests.


\begin{thebibliography}{10}
\expandafter\ifx\csname url\endcsname\relax
  \def\url#1{\texttt{#1}}\fi
\expandafter\ifx\csname urlprefix\endcsname\relax\def\urlprefix{URL }\fi
\providecommand{\bibinfo}[2]{#2}
\providecommand{\eprint}[2][]{\url{#2}}

\bibitem{Anderson1973}
\bibinfo{author}{Anderson, P.~W.}
\newblock \bibinfo{title}{{Resonating valence bonds: A new kind of insulator?}}
\newblock \emph{\bibinfo{journal}{Mater. Res. Bull.}}
  \textbf{\bibinfo{volume}{8}}, \bibinfo{pages}{153}
  (\bibinfo{year}{1973}).

 \bibitem{Pytte1974}
\bibinfo{author}{Pytte, E.}
\newblock \bibinfo{title}{{Peierls instability in Heisenberg chains}}.
\newblock \emph{\bibinfo{journal}{Phys. Rev. B}}
  \textbf{\bibinfo{volume}{10}}, \bibinfo{pages}{4637}
  (\bibinfo{year}{1974}).

\bibitem{Bray1975}
\bibinfo{author}{Bray, J.~W.} \emph{et~al.}
\newblock \bibinfo{title}{{Observation of a Spin-Peierls Transition in a Heisenberg Antiferromagnetic Linear-Chain System}}.
\newblock \emph{\bibinfo{journal}{Phys. Rev. Lett.}}
  \textbf{\bibinfo{volume}{35}}, \bibinfo{pages}{744}
  (\bibinfo{year}{1975}).

\bibitem{Jacobs1976}
\bibinfo{author}{Jacobs, I.~S.} \emph{et~al.}
\newblock \bibinfo{title}{{Spin-Peierls transitions in magnetic donor-acceptor compounds of tetrathiafulvalene (TTF) with bisdithiolene metal complexes}}.
\newblock \emph{\bibinfo{journal}{Phys. Rev. B}}
  \textbf{\bibinfo{volume}{14}}, \bibinfo{pages}{3036}
  (\bibinfo{year}{1976}).

\bibitem{Huizinga1979}
\bibinfo{author}{Huizinga, S.} \emph{et~al.}
\newblock \bibinfo{title}{{Spin- Peierls transition in N-methyl-N-ethyl-morpholinium- ditetracyanoquinodimethanide [MEM-(TCNQ)$_{2}$]}}.
\newblock \emph{\bibinfo{journal}{Phys. Rev. B}}
  \textbf{\bibinfo{volume}{19}}, \bibinfo{pages}{4723}
  (\bibinfo{year}{1979}).

  \bibitem{Hase1993}
\bibinfo{author}{Hase, M.} , \bibinfo{author}{Terasaki, I.}
   \& \bibinfo{author}{Uchinokura, K.}
\newblock \bibinfo{title}{{Observation of the spin-Peierls transition in linear Cu$^{2+}$ (spin-1/2) chains in an inorganic compound CuGeO$_{3}$}}.
\newblock \emph{\bibinfo{journal}{Phys. Rev. Lett.}}
  \textbf{\bibinfo{volume}{70}}, \bibinfo{pages}{3651}
  (\bibinfo{year}{1993}).

  \bibitem{Seidel2003}
\bibinfo{author}{Seidel, A.} \emph{et~al.}
\newblock \bibinfo{title}{{S=1/2 chains and spin-Peierls transition in TiOCl}}.
\newblock \emph{\bibinfo{journal}{Phys. Rev. B}}
  \textbf{\bibinfo{volume}{67}}, \bibinfo{pages}{020405(R)} (\bibinfo{year}{2003}).

  \bibitem{Riera1995}
\bibinfo{author}{Riera, J.} \& \bibinfo{author}{Dobry, A.}
\newblock \bibinfo{title}{{Magnetic susceptibility in the spin-Peierls system CuGeO$_{3}$}}.
\newblock \emph{\bibinfo{journal}{Phys. Rev. B}}
  \textbf{\bibinfo{volume}{51}}, \bibinfo{pages}{16098} (\bibinfo{year}{1996}).

\bibitem{Abel2007}
\bibinfo{author}{Abel, E.~T.} \emph{et~al.}
\newblock \bibinfo{title}{{X-ray scattering study of the spin-Peierls transition and soft phonon behavior in TiOCl}}.
\newblock \emph{\bibinfo{journal}{Phys. Rev. B}}
  \textbf{\bibinfo{volume}{76}}, \bibinfo{pages}{214304}
  (\bibinfo{year}{2007}).

 \bibitem{Shaz2005}
\bibinfo{author}{Shaz, M.} \emph{et~al.}
\newblock \bibinfo{title}{{Spin-Peierls transition in TiOCl}}.
\newblock \emph{\bibinfo{journal}{Phys. Rev. B}}
  \textbf{\bibinfo{volume}{71}}, \bibinfo{pages}{100405(R)} (\bibinfo{year}{2005}).

 \bibitem{Ruckamp2005}
\bibinfo{author}{R$\ddot{u}$ckamp, M.} \emph{et~al.}
\newblock \bibinfo{title}{{Zero-Field Incommensurate Spin-Peierls Phase with Interchain Frustration in TiOCl}}.
\newblock \emph{\bibinfo{journal}{Phys. Rev. Lett.}}
  \textbf{\bibinfo{volume}{95}}, \bibinfo{pages}{097203} (\bibinfo{year}{2005}).

 \bibitem{Krimmel2006}
\bibinfo{author}{Krimel, A.} \emph{et~al.}
\newblock \bibinfo{title}{{Incommensurate structure of the spin-Peierls compound TiOCl in zero and finite magnetic fields}}.
\newblock \emph{\bibinfo{journal}{Phys. Rev. B}}
  \textbf{\bibinfo{volume}{73}}, \bibinfo{pages}{172413} (\bibinfo{year}{2006}).

\bibitem{Schonleber2006}
\bibinfo{author}{Sch$\ddot{o}$nleber, S.} \emph{et~al.}
\newblock \bibinfo{title}{{Structure of the incommensurate phase of the quantum magnet TiOCl}}.
\newblock \emph{\bibinfo{journal}{Phys. Rev. B}}
  \textbf{\bibinfo{volume}{73}}, \bibinfo{pages}{214410} (\bibinfo{year}{2006}).

 \bibitem{Imai2003}
\bibinfo{author}{Imai, T.} \emph{et~al.}
\newblock \bibinfo{title}{{Novel Spin-Gap Behavior in Layered S=1/2 Quantum Spin System TiOCl}}.
 Preprint at \url{https://arxiv.org/abs/cond-mat/0301425v1} (\bibinfo{year}{2003}).

 \bibitem{Caimi2004}
\bibinfo{author}{Caimi, G.} \emph{et~al.}
\newblock \bibinfo{title}{{Infrared optical properties of the spin-1/2 quantum magnet TiOCl}}.
\newblock \emph{\bibinfo{journal}{Phys. Rev. B}}
  \textbf{\bibinfo{volume}{69}}, \bibinfo{pages}{125108} (\bibinfo{year}{2004}).

 \bibitem{Lemmens2004}
\bibinfo{author}{Clancy, J.~P.} \emph{et~al.}
\newblock \bibinfo{title}{{Giant phonon softening in the pseudogap phase of the quantum spin system TiOCl}}.
\newblock \emph{\bibinfo{journal}{Phys. Rev. B}}
  \textbf{\bibinfo{volume}{70}}, \bibinfo{pages}{134429} (\bibinfo{year}{2004}).

 \bibitem{Hemberger2005}
\bibinfo{author}{Hemberger, J.} \emph{et~al.}
\newblock \bibinfo{title}{{Heat capacity of the quantum magnet TiOCl}}.
\newblock \emph{\bibinfo{journal}{Phys. Rev. B}}
  \textbf{\bibinfo{volume}{72}}, \bibinfo{pages}{012420} (\bibinfo{year}{2005}).

 \bibitem{Clancy2007}
\bibinfo{author}{Clancy, J.~P.} \emph{et~al.}
\newblock \bibinfo{title}{{Commensurate fluctuations in the pseudogap and incommensurate spin-Peierls phases of TiOCl}}.
\newblock \emph{\bibinfo{journal}{Phys. Rev. B}}
  \textbf{\bibinfo{volume}{75}}, \bibinfo{pages}{100401(R)} (\bibinfo{year}{2007}).

 \bibitem{Beynon1993}
\bibinfo{author}{Beynon, R.~J.} \& \bibinfo{author}{Wilson, J.~A.}
\newblock \bibinfo{title}{{TiOCl, TiOBr-are these RVB d$^{1}$, S=1/2 materials? The results of scandium substitution set in the context of other S=1/2 systems of current interest for high-temperature superconductivity and the metal-insulator transition}}.
\newblock \emph{\bibinfo{journal}{J. Phys.: Condens. Matter}}
  \textbf{\bibinfo{volume}{5}}, \bibinfo{pages}{1983} (\bibinfo{year}{1993}).

 \bibitem{Craco2006}
\bibinfo{author}{Craco, L.}, \bibinfo{author}{Laad, M.~S.} \& \bibinfo{author}{M$\ddot{u}$ller-Hartmann, E.}
\newblock \bibinfo{title}{{Metallizing the Mott insulatror TiOCl by elctron doping}}.
\newblock \emph{\bibinfo{journal}{J. Phys.: Condens. Matter}}
  \textbf{\bibinfo{volume}{18}}, \bibinfo{pages}{10943} (\bibinfo{year}{2006}).

\bibitem{Kuntscher2010}
\bibinfo{author}{Kuntscher, C.~A.} \emph{et~al.}
\newblock \bibinfo{title}{{Possible metallization of the Mott insulators TiOCl and TiOBr: Effects of doping and external pressure}}.
\newblock \emph{\bibinfo{journal}{Eur. Phys. J. Special Topics}}
  \textbf{\bibinfo{volume}{180}}, \bibinfo{pages}{29}
  (\bibinfo{year}{2010}).

\bibitem{Zhang2010}
\bibinfo{author}{Zhang, Y.-Z.} \emph{et~al.}
\newblock \bibinfo{title}{{Can the Mott Insularor TiOCl be Metallized by Doping? A First-Principles Study}}.
\newblock \emph{\bibinfo{journal}{Phys. Rev. Lett.}}
  \textbf{\bibinfo{volume}{104}}, \bibinfo{pages}{146402}
  (\bibinfo{year}{2010}).

  \bibitem{Adachi2000}
\bibinfo{author}{Adachi, T.} \emph{et~al.}
\newblock \bibinfo{title}{{Superconducting Transition of (TMTTF)$_{2}$PF$_{6}$ above 50 kbar [TMTTF = Tetramethyltetrathiafulvalene]}}.
\newblock \emph{\bibinfo{journal}{J. Am. Chem. Soc.}}
  \textbf{\bibinfo{volume}{122}}, \bibinfo{pages}{3238}
  (\bibinfo{year}{2000}).

\bibitem{Jaccard2001}
\bibinfo{author}{Jaccard, D.} \emph{et~al.}
\newblock \bibinfo{title}{{From spin-Peierls to superconductivity: (TMTTF)$_{2}$PF$_{6}$ under high pressure}}.
\newblock \emph{\bibinfo{journal}{J. Phys.: Condens. Matter}}
  \textbf{\bibinfo{volume}{13}}, \bibinfo{pages}{L89}
  (\bibinfo{year}{2001}).

  \bibitem{Kuntscher2006}
\bibinfo{author}{Kuntscher, C.~A.} \emph{et~al.}
\newblock \bibinfo{title}{{Possible pressure-induced insulator-to-metal transition in low-dimensional TiOCl}}.
\newblock \emph{\bibinfo{journal}{Phys. Rev. B}}
  \textbf{\bibinfo{volume}{74}}, \bibinfo{pages}{184402}
  (\bibinfo{year}{2006}).

\bibitem{Kuntscher2007}
\bibinfo{author}{Kuntscher, C.~A.} \emph{et~al.}
\newblock \bibinfo{title}{{Pressure-induced metallization and structural phase transition of the Mott-Hubbard insulator TiOBr}}.
\newblock \emph{\bibinfo{journal}{Phys. Rev. B}}
  \textbf{\bibinfo{volume}{76}}, \bibinfo{pages}{241101(R)}
  (\bibinfo{year}{2007}).

\bibitem{Kuntscher2008}
\bibinfo{author}{Kuntscher, C.~A.} \emph{et~al.}
\newblock \bibinfo{title}{{Mott-Hubbard gap closure and structural phase transition in the oxyhalides TiOBr and TiOCl under pressure}}.
\newblock \emph{\bibinfo{journal}{Phys. Rev. B}}
  \textbf{\bibinfo{volume}{78}}, \bibinfo{pages}{035106}
  (\bibinfo{year}{2008}).

\bibitem{Forthaus2008}
\bibinfo{author}{Forthaus, M.~K.} \emph{et~al.}
\newblock \bibinfo{title}{{Effect of pressure on the elctrical transort and structure of TioCl}}.
\newblock \emph{\bibinfo{journal}{Phys. Rev. B}}
  \textbf{\bibinfo{volume}{77}}, \bibinfo{pages}{165121}
  (\bibinfo{year}{2008}).

  \bibitem{Blanco2009}
\bibinfo{author}{Blanco-Canosa, S.} \emph{et~al.}
   \newblock \bibinfo{title}{{Enhanced dimerization of TiOCl under pressure: spin-Peierls to Peierls transition}}.
\newblock \emph{\bibinfo{journal}{Phys. Rev. B}}
  \textbf{\bibinfo{volume}{102}}, \bibinfo{pages}{056406} (\bibinfo{year}{2009}).

\bibitem{Ebad2010}
\bibinfo{author}{Ebad-Allah, J.} \emph{et~al.}
\newblock \bibinfo{title}{{Two pressure-induced structural phase transitions in TiOCl}}.
\newblock \emph{\bibinfo{journal}{Phys. Rev. B}}
  \textbf{\bibinfo{volume}{82}}, \bibinfo{pages}{134117}
  (\bibinfo{year}{2010}).

\bibitem{Prodi2010}
\bibinfo{author}{Prodi, A.} \emph{et~al.}
   \newblock \bibinfo{title}{{Pressure-induced spin-Peierls to incommensurate charge-density-wave transition in the ground stae of TiOCl}}.
\newblock \emph{\bibinfo{journal}{Phys. Rev. B}}
  \textbf{\bibinfo{volume}{81}}, \bibinfo{pages}{201102(R)} (\bibinfo{year}{2010}).

\bibitem{daSilvaNeto2014}
\bibinfo{author}{da Silva Neto, E.~H.} \emph{et~al.}
   \newblock \bibinfo{title}{{Ubiquitous Interplay Between Charge Ordering and High-Temperature Superconductivity in Cuprates}}.
\newblock \emph{\bibinfo{journal}{Science}}
  \textbf{\bibinfo{volume}{343}}, \bibinfo{pages}{393} (\bibinfo{year}{2014}).

 \bibitem{Kataev2003}
\bibinfo{author}{Kataev, V.} \emph{et~al.}
\newblock \bibinfo{title}{{Orbital order in the low-dimensional quantum spin system TiOCl probed by ESR}}.
\newblock \emph{\bibinfo{journal}{Phys. Rev. B}}
  \textbf{\bibinfo{volume}{68}}, \bibinfo{pages}{140405(R)} (\bibinfo{year}{2003}).

  \bibitem{Igor2009}
\bibinfo{author}{Walter, A.C.} \emph{et~al.}
\newblock \bibinfo{title}{{Effect of covalent bonding on magnetism and the missing neutron intensity in copper oxide compounds}}.
\newblock \emph{\bibinfo{journal}{Nat Phys}}
  \textbf{\bibinfo{volume}{5}}, \bibinfo{pages}{867--872} (\bibinfo{year}{2009}).

\bibitem{Yu2004}
\bibinfo{author}{Yu, W.} \emph{et~al.}
\newblock \bibinfo{title}{{Electron-lattice coupling and broken symmetries of the molecular salt (TMTTF)$_{2}$PF$_{6}$}}.
\newblock \emph{\bibinfo{journal}{Phys. Rev. B}}
  \textbf{\bibinfo{volume}{70}}, \bibinfo{pages}{121101(R)}
  (\bibinfo{year}{2004}).

\bibitem{Dressel2007}
\bibinfo{author}{Dressel, M.}
\newblock \bibinfo{title}{{Ordering phenomena in quasi-one-dimensional organic conductors}}.
\newblock \emph{\bibinfo{journal}{Naturwissenschaften}}
  \textbf{\bibinfo{volume}{94}}, \bibinfo{pages}{527}
  (\bibinfo{year}{2007}).

\bibitem{Jerome1991}
\bibinfo{author}{J$\acute{e}$rome, D.}
\newblock \bibinfo{title}{{The physics of organic superconducotrs}}.
\newblock \emph{\bibinfo{journal}{Science}}
  \textbf{\bibinfo{volume}{252}}, \bibinfo{pages}{1509}
  (\bibinfo{year}{1991}).

\bibitem{Tajima2002}
\bibinfo{author}{Tajima, N.} \emph{et~al.}
\newblock \bibinfo{title}{{Effects of uniaxial strain on transport properties of organic conductor alpha-(BEDT-TTF)$_{2}$I$_{3}$ and discovery of superconductivity}}.
\newblock \emph{\bibinfo{journal}{J. Phys. Soc. Jpn.}}
  \textbf{\bibinfo{volume}{71}}, \bibinfo{pages}{1832} (\bibinfo{year}{2002}).

\bibitem{Schafer1958}
\bibinfo{author}{Sch$\ddot{a}$fer, H.}, \bibinfo{author}{Wartenpfuhl, F.}
   \& \bibinfo{author}{Weise, E.}
\newblock \bibinfo{title}{{$\ddot{U}$ber Titanchloride. V. Titan(III)-oxychlorid}}.
\newblock \emph{\bibinfo{journal}{Z. Anorg. Allg. Chem.}}
  \textbf{\bibinfo{volume}{295}}, \bibinfo{pages}{268}
  (\bibinfo{year}{1958}).

\bibitem{Tateiwa2009}
\bibinfo{author}{Tateiwa, N.} \& \bibinfo{author}{Haga, Y.}
\newblock \bibinfo{title}{{Evaluations of pressure-transmitting media for cryogenic experiments with diamond anvil cell}}.
\newblock \emph{\bibinfo{journal}{Rev. Sci. Instrum.}}
  \textbf{\bibinfo{volume}{80}}, \bibinfo{pages}{123901} (\bibinfo{year}{2009}).

\bibitem{Yejun2010}
\bibinfo{author}{Feng, Y.} \emph{et~al.}
\newblock \bibinfo{title}{{High-pressure techniques for condensed matter physics at low temperature}}.
\newblock \emph{\bibinfo{journal}{Rev. Sci. Instrum.}}
  \textbf{\bibinfo{volume}{81}}, \bibinfo{pages}{041301} (\bibinfo{year}{2010}).

\bibitem{Dewaele2006}
\bibinfo{author}{Dewaele, A.} , \bibinfo{author}{Loubeyre, P.}
   \& \bibinfo{author}{Andr$\acute{e}$, R.}
\newblock \bibinfo{title}{{An x-ray topographic study of diamond anvils: Correlation between defects and helium
diffusion}}.
\newblock \emph{\bibinfo{journal}{J. Appl. Phys.}}
  \textbf{\bibinfo{volume}{99}}, \bibinfo{pages}{104906} (\bibinfo{year}{2006}).

\end{thebibliography}
\end{document}